


\font\bigbold=cmbx12
\font\ninerm=cmr9
\font\eightrm=cmr8
\font\sixrm=cmr6
\font\fiverm=cmr5
\font\ninebf=cmbx9
\font\eightbf=cmbx8
\font\sixbf=cmbx6
\font\fivebf=cmbx5
\font\ninei=cmmi9  \skewchar\ninei='177
\font\eighti=cmmi8  \skewchar\eighti='177
\font\sixi=cmmi6    \skewchar\sixi='177
\font\fivei=cmmi5
\font\ninesy=cmsy9 \skewchar\ninesy='60
\font\eightsy=cmsy8 \skewchar\eightsy='60
\font\sixsy=cmsy6   \skewchar\sixsy='60
\font\fivesy=cmsy5
\font\nineit=cmti9
\font\eightit=cmti8
\font\ninesl=cmsl9
\font\eightsl=cmsl8
\font\ninett=cmtt9
\font\eighttt=cmtt8
\font\tenfrak=eufm10
\font\ninefrak=eufm9
\font\eightfrak=eufm8
\font\sevenfrak=eufm7
\font\fivefrak=eufm5
\font\tenbb=msbm10
\font\ninebb=msbm9
\font\eightbb=msbm8
\font\sevenbb=msbm7
\font\fivebb=msbm5
\font\tensmc=cmcsc10


\newfam\bbfam
\textfont\bbfam=\tenbb
\scriptfont\bbfam=\sevenbb
\scriptscriptfont\bbfam=\fivebb
\def\Bbb{\fam\bbfam}

\newfam\frakfam
\textfont\frakfam=\tenfrak
\scriptfont\frakfam=\sevenfrak
\scriptscriptfont\frakfam=\fivefrak
\def\frak{\fam\frakfam}

\def\smc{\tensmc}


\def\eightpoint{%
\textfont0=\eightrm   \scriptfont0=\sixrm
\scriptscriptfont0=\fiverm  \def\rm{\fam0\eightrm}%
\textfont1=\eighti   \scriptfont1=\sixi
\scriptscriptfont1=\fivei  \def\oldstyle{\fam1\eighti}%
\textfont2=\eightsy   \scriptfont2=\sixsy
\scriptscriptfont2=\fivesy
\textfont\itfam=\eightit  \def\it{\fam\itfam\eightit}%
\textfont\slfam=\eightsl  \def\sl{\fam\slfam\eightsl}%
\textfont\ttfam=\eighttt  \def\tt{\fam\ttfam\eighttt}%
\textfont\frakfam=\eightfrak \def\frak{\fam\frakfam\eightfrak}%
\textfont\bbfam=\eightbb  \def\Bbb{\fam\bbfam\eightbb}%
\textfont\bffam=\eightbf   \scriptfont\bffam=\sixbf
\scriptscriptfont\bffam=\fivebf  \def\bf{\fam\bffam\eightbf}%
\abovedisplayskip=9pt plus 2pt minus 6pt
\belowdisplayskip=\abovedisplayskip
\abovedisplayshortskip=0pt plus 2pt
\belowdisplayshortskip=5pt plus2pt minus 3pt
\smallskipamount=2pt plus 1pt minus 1pt
\medskipamount=4pt plus 2pt minus 2pt
\bigskipamount=9pt plus4pt minus 4pt
\setbox\strutbox=\hbox{\vrule height 7pt depth 2pt width 0pt}%
\normalbaselineskip=9pt \normalbaselines
\rm}


\def\ninepoint{%
\textfont0=\ninerm   \scriptfont0=\sixrm
\scriptscriptfont0=\fiverm  \def\rm{\fam0\ninerm}%
\textfont1=\ninei   \scriptfont1=\sixi
\scriptscriptfont1=\fivei  \def\oldstyle{\fam1\ninei}%
\textfont2=\ninesy   \scriptfont2=\sixsy
\scriptscriptfont2=\fivesy
\textfont\itfam=\nineit  \def\it{\fam\itfam\nineit}%
\textfont\slfam=\ninesl  \def\sl{\fam\slfam\ninesl}%
\textfont\ttfam=\ninett  \def\tt{\fam\ttfam\ninett}%
\textfont\frakfam=\ninefrak \def\frak{\fam\frakfam\ninefrak}%
\textfont\bbfam=\ninebb  \def\Bbb{\fam\bbfam\ninebb}%
\textfont\bffam=\ninebf   \scriptfont\bffam=\sixbf
\scriptscriptfont\bffam=\fivebf  \def\bf{\fam\bffam\ninebf}%
\abovedisplayskip=10pt plus 2pt minus 6pt
\belowdisplayskip=\abovedisplayskip
\abovedisplayshortskip=0pt plus 2pt
\belowdisplayshortskip=5pt plus2pt minus 3pt
\smallskipamount=2pt plus 1pt minus 1pt
\medskipamount=4pt plus 2pt minus 2pt
\bigskipamount=10pt plus4pt minus 4pt
\setbox\strutbox=\hbox{\vrule height 7pt depth 2pt width 0pt}%
\normalbaselineskip=10pt \normalbaselines
\rm}


\def\pagewidth#1{\hsize= #1}
\def\pageheight#1{\vsize= #1}
\def\hcorrection#1{\advance\hoffset by #1}
\def\vcorrection#1{\advance\voffset by #1}

\newif\iftitlepage   \titlepagetrue               
\newtoks\titlepagefoot     \titlepagefoot={\hfil} 
\newtoks\otherpagesfoot    \otherpagesfoot={\hfil\tenrm\folio\hfil}
\footline={\iftitlepage\the\titlepagefoot\global\titlepagefalse
           \else\the\otherpagesfoot\fi}

\font\extra=cmss10 scaled \magstep0
\setbox1 = \hbox{{{\extra R}}}
\setbox2 = \hbox{{{\extra I}}}
\setbox3 = \hbox{{{\extra C}}}
\setbox4 = \hbox{{{\extra Z}}}
\setbox5 = \hbox{{{\extra N}}}

\def\RRR{{{\extra R}}\hskip-\wd1\hskip2.0 true pt{{\extra I}}\hskip-\wd2
\hskip-2.0 true pt\hskip\wd1}
\def\Real{\hbox{{\extra\RRR}}}    


\def\ZZZ{{{\extra Z}}\hskip-\wd4\hskip 2.5 true pt{{\extra Z}}}
\def\Zed{\hbox{{\extra\ZZZ}}}       



\def\g{{\frak g}}
\def\h{{\frak h}}
\def\r{{\frak r}}

\def\dg{{\cal D}g}
\def\dh{{\cal D}h}
\def\dq{{\cal D}q}

\def\Z{{\Zed}}
\def\R{{\Real}}
\def\a{\alpha}

\def\s{\sigma}

\def\ket#1{|#1\rangle}
\def\bra#1{\langle#1|}
\def\tr{\hbox{tr}}
\def\la{\langle}
\def\ra{\rangle}
\def\pr{\pi}


\def\abstract#1{{\parindent=30pt\narrower\noindent\ninepoint\openup
2pt #1\par}}


\newcount\notenumber\notenumber=1
\def\note#1
{\unskip\footnote{$^{\the\notenumber}$}
{\eightpoint\openup 1pt #1}
\global\advance\notenumber by 1}


\def\frac#1#2{{#1\over#2}}

\def\({\left(}
\def\){\right)}
\def\<{\langle}
\def\>{\rangle}

\def\pmb#1{\setbox0=\hbox{$#1$}%
   \kern-.025em\copy0\kern-\wd0
   \kern.05em\copy0\kern-\wd0
   \kern-.025em\raise.0433em\box0 }


\global\newcount\secno \global\secno=0
\global\newcount\meqno \global\meqno=1
\global\newcount\appno \global\appno=0
\newwrite\eqmac
\def\romappno{\ifcase\appno\or A\or B\or C\or D\or E\or F\or G\or H
\or I\or J\or K\or L\or M\or N\or O\or P\or Q\or R\or S\or T\or U\or
V\or W\or X\or Y\or Z\fi}
\def\eqn#1{
        \ifnum\secno>0
            \eqno(\the\secno.\the\meqno)\xdef#1{\the\secno.\the\meqno}
          \else\ifnum\appno>0
            \eqno({\rm\romappno}.\the\meqno)\xdef#1{{\rm\romappno}.\the\meqno}
          \else
            \eqno(\the\meqno)\xdef#1{\the\meqno}
          \fi
        \fi
\global\advance\meqno by1 }


\global\newcount\refno
\global\refno=1 \newwrite\reffile
\newwrite\refmac
\newlinechar=`\^^J
\def\ref#1#2{\the\refno\nref#1{#2}}
\def\nref#1#2{\xdef#1{\the\refno}
\ifnum\refno=1\immediate\openout\reffile=refs.tmp\fi
\immediate\write\reffile{
     \noexpand\item{[\noexpand#1]\ }#2\noexpand\nobreak.}
     \immediate\write\refmac{\def\noexpand#1{\the\refno}}
   \global\advance\refno by1}
\def\semi{;\hfil\noexpand\break ^^J}
\def\nl{\hfil\noexpand\break ^^J}
\def\refn#1#2{\nref#1{#2}}

\def\ann#1#2#3{{\it Ann. Phys.} {\bf {#1}} (19{#2}) #3}
\def\cmp#1#2#3{{\it Commun. Math. Phys.} {\bf {#1}} (19{#2}) #3}
\def\jgp#1#2#3{{\it J. Geom. Phys.} {\bf {#1}} (19{#2}) #3}
\def\jmp#1#2#3{{\it J. Math. Phys.} {\bf {#1}} (19{#2}) #3}
\def\jpA#1#2#3{{\it J.  Phys.} {\bf A{#1}} (19{#2}) #3} 
 
\def\mplA#1#2#3{{\it Mod.  Phys.  Lett.} {\bf A{#1}} (19{#2}) #3} 
\def\nc#1#2#3{{\it Nuovo Cimento} {\bf {#1}D} (19{#2}) #3} 
\def\np#1#2#3{{\it Nucl.  Phys.} {\bf B{#1}} (19{#2}) #3} 
\def\pl#1#2#3{{\it Phys.  Lett.} {\bf {#1}B} (19{#2}) #3} 
\def\pr#1#2#3{{\it Phys.  Rev.} {\bf {#1}} (19{#2}) #3} 
 
\def\prD#1#2#3{{\it Phys.  Rev.} {\bf D{#1}} (19{#2}) #3}


{
\refn\LDW 
{M.G.G.  Laidlaw and C.M.  DeWitt, \prD{3}{71}{1375}}

\refn\Sch
{L.S. Schulman, \pr{176}{68}{1558}}

\refn\Schulman
{L.S. Schulman, \lq\lq Techniques and Applications of
Path Integration\rq\rq, Willey, New York, 1981}

\refn\Dowker 
{J.S.  Dowker, \jpA{5}{72}{936}}

\refn\HMS 
{P.A.  Horvathy, G.  Morandi and E.C.G.  Sudarshan,
\nc{11}{89}{201}}

\refn\Mackey
{G.W. Mackey, \lq\lq Induced Representations of Groups
and Quantum Mechanics\rq\rq,\nl Benjamin, New York, 1969}

\refn\Isham
{C.J. Isham, in \lq\lq Relativity, Groups and Topology
II\rq\rq, (B.S. DeWitt and R. Stora, Eds)\nl North-Holland,
Amsterdam, 1984}

\refn\LL
{N.P. Landsman and N. Linden, \np{365}{91}{121}}

\refn\KN
{S. Kobayashi and K. Nomizu, \lq\lq Foundations of Differential
Geometry\rq\rq\
Vols.I,II (Interscience, New York, 1969)}

\refn\BB
{F.A. Bais and P. Batenburg, \np{269}{86}{363}}

\refn\OK
{Y. Ohnuki and S. Kitakado, \mplA{7}{92}{2477}; \jmp{34}{93}{2827}}

\refn\TT
{S. Tanimura and I. Tsutsui, \mplA{34}{95}{2607}}

\refn\letter
{D. McMullan and I. Tsutsui, \pl{320}{94}{287}}

\refn\MT
{D. McMullan and I. Tsutsui, \ann{237}{95}{269}}

\refn\Dirac
{P.A.M. Dirac, \lq\lq Lectures on Quantum
Mechanics\rq\rq, Yeshiva, New York, 1964}

\refn\Klauder
{J.R. Klauder and B.-S. Skagerstam (Eds.), 
\lq\lq Coherent States\rq\rq,
World Scientific, Singapore, 1985}

\refn\AB
{Y. Aharonov and D. Bohm, \pr{115}{59}{485}}

\refn\Humphreys
{S.E. Humphreys, \lq\lq Introduction to Lie Algebras
and Representation Theory\rq\rq,\nl
Springer-Verlag, Berlin, 1972}

\refn\AM
{R. Abraham and J.E. Marsden, \lq\lq Foundations of Mechanics\rq\rq,
Second Edition,\nl Benjamin/Cummings, Reading, 1978}

\refn\Perelomov
{A.M. Perelomov, \lq\lq Integrable Systems of
Classical Mechanics and Lie Algebras\rq\rq,\nl
Birkh\"auser Verlag, Basel, 1990}

\refn\Wong
{S.K. Wong, {\it Nuovo Cimento} {\bf 65A} (1970), 689}

\refn\DP
{D.I. Diakonov and V.Yu. Petrov, \pl{224}{89}{131}}

\refn\Montgomery
{R. Montgomery, \cmp{128}{90}{565}}

\refn\Robson
{M. A. Robson, \pl{335}{94}{383}; \jgp{19}{96}{207}}
 
\refn\LMT
{P. L\'{e}vay, D. McMullan and I. Tsutsui, \jmp{37}{96}{625}}

\refn\Barut
{A.O. Barut and R. Raczka, \lq\lq Theory of Group
Representations and Applications\rq\rq,\nl
Polish Scientific Publishers, Warszawa, 1977}

}


\pageheight{23cm}
\pagewidth{15.7cm}
\hcorrection{-1mm}
\magnification= \magstep1
\def\bsk{%
\baselineskip=14.5pt plus 1pt minus 1pt}
\parskip=5pt plus 1pt minus 1pt
\tolerance 8000



\null
\rightline{KU-AMP-96009}
\rightline{INS-Rep.-1157} 
\rightline{September 1996} 
\smallskip
\vfill 
{\baselineskip=18pt
\centerline{\bigbold Inequivalent Quantizations and Holonomy Factor} 
\centerline{\bigbold from the Path-Integral Approach}

\vskip 30pt
\centerline{\smc Shogo Tanimura}
\vskip 5pt
{
\baselineskip=13pt
\centerline{Department of Applied Mathematics and Physics}
\centerline{Graduate School of Engineering}
\centerline{Kyoto University}
\centerline{Kyoto 606-01} 
\centerline{Japan}
\centerline{(e-mail: tanimura@kuamp.kyoto-u.ac.jp)}
}
\vskip 10pt
\centerline{\smc and}
\vskip 8pt
\centerline{\smc Izumi Tsutsui}
\vskip 5pt
{
\baselineskip=13pt
\centerline{Institute for Nuclear Study}
\centerline{University of Tokyo}
\centerline{Midori-cho, Tanashi-shi, Tokyo 188}
\centerline{Japan}
\centerline{(e-mail: tsutsui@insth1.ins.u-tokyo.ac.jp)}
}
\vskip 60pt
\abstract{%
{\bf Abstract.}\quad 
A path-integral quantization on a homogeneous space $G/H$ 
is proposed based on the guiding principle \lq first lift to $G$ and 
then project to $G/H$'.  It is then shown that this principle 
gives a simple procedure to obtain the inequivalent quantizations 
(superselection sectors) along with the holonomy factor (induced gauge 
field) found earlier by algebraic approaches.  
We also prove that the resulting
matrix-valued path-integral is physically equivalent to 
the scalar-valued path-integral derived in the Dirac approach, 
and thereby present a unified viewpoint to discuss the basic features of 
quantizing on $G/H$ obtained in various approaches so far.
}

\vfill\eject

\bsk

\noindent \secno=1 \meqno=1

\centerline
{\bf 1.  Introduction} 
\medskip

Nearly a quarter of a century ago, Laidlaw and DeWitt [\LDW]
studied, generalizing Schulman's idea [\Sch] (see also [\Schulman]),
the path-integral quantization on a multiply connected
configuration space $Q$,
and established the, by now well-known,
path-integral formula for a transition amplitude given by
summing up paths with weight 
factors characterized by the homotopy class of the path.
Subsequently, Dowker [\Dowker]
reconsidered their argument using the concept of 
covering space to provide a convenient and geometric 
framework to deal with the paths of different
homotopy classes.   In this covering space
construction one considers the universal covering
space $\bar Q$ of $Q$, for which $Q = \bar Q / \Gamma$
with $\Gamma$ being the discrete group of isometries
of $\bar Q$ which is isomorphic to the fundamental
group of the space, $\pi_1(Q)$.  
Thus, if $\bar q_0 \in \bar Q$ is a representative point then 
every other point in $\bar Q$ which reduces to a point $q 
\in Q$ under the covering projection $\pi:\, \bar Q \rightarrow Q$ 
can be written as $\bar q = \bar q_0 \gamma$ using some $\gamma \in 
\Gamma$, that is, $\pi(\bar q_0 \gamma) = q$ for any $\gamma \in 
\Gamma$.  The path-integral formula for the propagator $K^Q(q', q; T)$ on 
$Q$ put forward in this construction then takes the form,
$$
K^Q(q', q; T) = \sum_{\gamma \in \Gamma} \rho(\gamma) \,
K^{\bar Q}(\bar q'\gamma, \bar q; T)\ .
\eqn\ldw
$$
A salient feature of this formula (\ldw) is that
the propagator on the multiply connected space
$Q$ is defined with the help of the
propagator $K^{\bar Q}(\bar q'\gamma, \bar q; T)
= \int_{\bar q}^{\bar q'\gamma} {\cal D} \bar q\, e^{iS_{\bar Q}(\bar q)}$
on the covering space $\bar Q$, where $\bar q$ and $\bar 
q'$ are chosen such that $\pi(\bar q) = q$ and $\pi(\bar q') = q'$.  
Another important point to note is that the weight factors 
$\rho(\gamma)$ form unitary irreducible representations
of the fundamental group\note{%
More precisely, the weight factors form unitary representations
of the homology group $H_1(Q, \Z)$,
{\it i.e.}, the \lq Abelianized' $\pi_1(Q)$; see [\HMS].}
$\pi_1(Q)$ [\LDW, \Dowker].  One therefore finds many 
{\it inequivalent quantizations} (superselection sectors) depending on 
which representation of $\Gamma$ one uses for $\rho(\gamma)$.  
In the 
standard \lq scalar quantum mechanics' where one uses scalar-valued 
(one-component) wave functions, 
the unitary representation must be one-dimensional, whereas 
it becomes multi-dimensional when, due to some internal symmetries, 
vector-valued (multi-component) wave functions are considered 
[\HMS].

Meanwhile, even prior to the above investigations, 
quantization on a configuration space of the form $Q = G/H$,
where $G$ is a Lie group and $H$ its subgroup, 
was initiated by Mackey [\Mackey] (see also [\Isham]) based
on the imprimitivity relations, which is a generalization of the
canonical commutation relations (Weyl relations).  
It was then found that such a 
homogeneous space $Q = G/H$ admits inequivalent quantizations 
which are labelled by the irreducible representations of $H$.  
Moreover, it was uncovered recently [\LL] in investigating the 
dynamical consequence of these inequivalent quantizations that 
different superselection sectors come equipped with a specific type of 
induced gauge field $A(q)$, called the {\it canonical connection} (or 
the $H$-connection) on the homogeneous space $G/H$, which is a 
solution of the Yang-Mills equations on $G/H$ [\KN, \BB].  (A similar 
result has been obtained for the case $Q = S^n$ in [\OK].) The 
approaches employed in the above analyses are based purely on algebraic 
constructions, but it is worthwhile to mention that 
the resulting matrix-valued path-integral is characterized
geometrically by the holonomy factor ${\cal P}\,
{\rm exp}(- \int_0^T dt\, A)$ associated with the induced
connection.
Interestingly, one can also show that the propagator on $G/H$ admits a 
formula similar to (\ldw) where now $\bar Q$ and $\Gamma$ are replaced 
by $G$ and $H$, respectively [\LL].

In view of the similarity in the outcome of 
quantization between the two types of
configuration spaces, $Q = \bar Q/\Gamma$ and $Q = G/H$, we
are tempted to reverse the logic and set up a quantization procedure
for the latter type of spaces purely by the path-integral approach.   
In other words, we are interested in deriving all the known
results obtained by algebraic approaches 
for a homogeneous space $Q = G/H$ starting from a path-integral 
formula defined along the line of [\LDW, \Schulman, \Dowker] for a 
multiply connected space $Q = \bar Q/\Gamma$.  
The aim of 
this paper is to carry out this programme, by completing our earlier
attempt made in [\TT].  Indeed, we shall show 
that it is not only possible but also quite simple to do this, once 
we follow a geometrically motivated guiding principle which underlies
the covering space construction of the path-integral for a
multiply connected space.
Note however that
for $Q = \bar Q/ \Gamma$ the inequivalent quantizations arise only if
the fundamental group $\pi_1(Q)$ of the space is 
nontrivial whereas for $Q = G/H$ 
they arise even if it is trivial ($\pi_1(Q) = 0$),
although for both types of spaces
they are characterized by the
irreducible representations of the subgroup $\Gamma$ or $H$.
Thus the path-integral approach that we are going to 
discuss may be regarded as 
a generalization of the approach proposed for the 
quantization on $Q = \bar Q/ \Gamma$, but the important point 
is that this generalization is 
{\it necessary}
to reproduce the 
known results of quantization on $G/H$.
In fact, such a generalization was mentioned in [\Dowker] but 
dropped on the ground that it might have modified the dynamics, which we now 
know that it should do, due to the induced connection.

Another approach to 
quantizing on $G/H$ was developed more recently in [\letter, \MT], 
where the system of interest on $G/H$ is regarded as an effective 
constrained system on $G$ and quantized using the Dirac approach 
[\Dirac] to constrained systems.  One of the advantages of this 
picture is that one obtains a scalar-valued path-integral (at 
the expense of introducing extra degrees of freedom) which is easier 
to deal with than the matrix-valued one in the previous 
approaches.  We will show however that this scalar-valued 
path-integral can be derived from the matrix-valued path-integral 
using the technique of coherent state path-integrals [\Klauder], 
hoping that the path-integral approach presented here will
furnish a unified viewpoint to discuss the various features of 
quantizing on $G/H$ obtained by different approaches so far.

The plan of the paper is as follows: In Section 2 we set up the 
procedures of the (generalized) path-integral quantization based on 
the principle mentioned above, and show that the path-integral 
leads correctly to the inequivalent quantizations possessing
the holonomy factor (induced connection) in the required 
form.  Then the equivalence of the two path-integrals, scalar-valued 
and matrix-valued, is proven in Section 3.  Section 4 is devoted to the 
Conclusion 
and discussion, where a possible application and extension of the 
path-integral is also discussed.  An appendix is provided to collect
conventions used in the text.

\bigskip

\noindent \secno=2 \meqno=1

\centerline
{\bf 2. The path-integral on a homogeneous space $Q$}
\medskip

In this section we wish to 
show that 
the path-integral on a homogeneous space $Q$ describing
Mackey's inequivalent quantizations
can readily be
derived if we adopt the guiding principle that 
{\it we first lift to $G$ and then project to $G/H$},
where $G$ is a simply connected group for which $Q = G/H$,
allowing a phase factor in the path-integral.
This is basically the same idea used in the covering space
construction for the propagator on a multiply connected space,
but we first formulate it more precisely on 
a homogeneous space $G/H$,
and then prove that the matrix-valued path-integral
with the holonomy factor mentioned earlier
is a direct consequence of our
definition.  
The implications of the guiding principle will be discussed later.

\medskip
\noindent
{\bf 2.1. Definition}

In order to quantize a 
classical system whose configuration space $Q$ is
a homogeneous space, we shall adopt, according to
our guiding principle, 
the following three-step programme:

\item{(i)} 
{}Find a simply connected, semisimple group $G$ with which the 
configuration space is regarded as $Q = G/H$ with $H$ a closed 
subgroup of $G$.
\item{(ii)}
Choose a classical system on $G$ which is invariant 
under the action of $H$ such that its classical reduction
to $G/H$ gives the system on $Q$.
\item{(iii)} 
Define the path-integral on $Q = G/H$ by 
projecting the path-integral on $G$
down to $Q$, allowing 
a phase factor to appear.

To spell out the last step more explicitly,
let us consider the propagator on the lifted system of $G$,
$$
K^G ( g' , g ; T )
        =
        \int_g^{g'} \dg \, e^{i S_G(g)} \ ,
\eqn\piong
$$
which is invariant under
the right translation with respect to the subgroup $H$, 
$$
K^G ( g'h , g h ; T ) = K^G ( g' , g ; T )\ .
\eqn\hinv
$$
The measure $\dg$ 
in the path-integral (\piong) 
is formally a product of the (normalized) Haar measure $dg$ on $G$ 
which is (right) invariant, $d(gh) = dg$.  Thus
the invariance (\hinv) is guaranteed if the action 
$S_G(g)$ of the system is invariant under (time independent)
transformations
$S_G(gh) = S_G(g)$.  Equivalently, the potential
in the action
$$
 S_G(g) = \int_0^T dt
                \bigl(
                \frac{1}{2} || \dot g ||^2 - V ( g )
                \bigr) \ ,
\eqn\gaction
$$
which reduces to the potential on $Q$ upon projection, 
is assumed to be invariant $V(gh) = V(g)$. 
In (\gaction) the dot denotes the time derivative $\dot g = dg/dt$, and 
the norm $ || \cdot || $ is given by the invariant metric on $ G $, that 
is, $ || \dot{g} ||^2 := \tr ( g^{-1} \dot{g} )^2 $ with \lq $\tr$' 
being a matrix trace properly normalized in some irreducible 
representation (for our conventions used in this paper, see Appendix).

To furnish
necessary geometry to implement the lift and projection step,
we recall that, with 
$G/H$ viewed as the left cosets $\{gH\, \vert \, g \in G\}$, 
any element $g \in G$ admits the decomposition,
$$
g = \s(q)h\ ,
\eqn\candec
$$ 
where $\s(q) \in G$ is a map $G/H \rightarrow G$ and $h \in H$.  If we regard 
$G$ as a principal bundle $G(G/H, H)$ where $G/H$ is the base space 
and $H$ the fibre, then the map $\s(q)$ provides a section on the 
bundle.  With this section the projection map $\pi: G \rightarrow G/H$ 
given by $\pi(g):= gH$ satisfies $\pi(\s(q))=q$, thus yielding $\pi(g)= 
q$ when the decomposition (\candec) is used.  (Unless the bundle is 
trivial, the section $\s$ is defined only locally, and we need to 
introduce an open covering $ \{ D_\alpha \} $ of $ Q = \cup_\alpha 
D_\alpha $ on which local sections $ \{ \s_\alpha : D_\alpha \to G \} $ 
are given.  In this paper, however, for brevity we ignore the treatment 
necessary to deal with the locality of the sections; see [\TT].)

Given a path $q(t)$ in $Q$ with $q(0) = q$ and
$q(T) = q'$, 
the geometrical setting provided above allows us to 
lift it to the path $g(t) = \s(q(t)) h(t)$ in $G$ with
$h(0) = 1$ and $h(T) = h$.  We can then consider the propagation from $q$ 
to $q'$ on $Q = G/H$ by the corresponding lifted propagation from $\s(q)$ 
to $\s(q')h$ on $G$, where $h \in H$ represents the redundancy along the 
fibre (we need to consider the redundancy only at the final point $t = 
T$ because of the invariance (\hinv)).  Thus,
with the (normalized) Haar measure $dh$ on $H$ we define the 
path-integral on $G/H$ by the following form:
$$
K^{G/H}(q', q; T) := 
\int_H dh \, \rho(h) \, K^G(\s(q')h, \s(q); T) \ .
\eqn\pidef
$$ 
The integration over $h \in H$ is performed in order to 
implement the projection
from $G$ to $Q$ by summing
up the final points $\s(q')h$.
The point to be noted here is that 
in this projection we allow an $h$-dependent weight 
factor $\rho(h)$ to appear, in analogy with the case
of multiply connected spaces (\ldw).    

The weight factor is not 
entirely arbitrary but subject to the condition that the propagator in 
(\pidef) must fulfill the composition law,
$$
K^{G/H}(q', q; T) 
= \int_Q dq'' \, K^{G/H}(q', q''; T-t) \, K^{G/H}(q'', q; t) \ ,
\eqn\compgh
$$
where $dq''$ is the $G$-invariant measure on $Q = G/H$ induced
from the measure $dg$ through the decomposition (\candec), {\it i.e.},
$$
\int_G dg\, f(g) = \int_Q dq \int_H dh \, f(\s(q) h)\ ,
\eqn\decint
$$
for a function $f(g)$ on $G$.
Using the composition law for the propagator
$K^G(g', g; T)$ on $G$, one finds that (\compgh) holds if 
$$
\rho (h) \, \rho(h') = \rho(h h')\ , \qquad \hbox{for}
\quad h, \, h' \in H\ .
\eqn\rep
$$
This property (\rep) implies that the
weight factor furnishes 
a representation of the group $H$.
Thus, unless the representation is trivial and one-dimensional,
the wave functions on which the propagator (\pidef) acts are
vector-valued, and the Hilbert space to which these wave functions
belong is specified by the representation used.  For definiteness, 
we consider for $\rho(h)$ 
a highest weight
representation labeled by the highest weight $\chi$ (see Appendix),
$$
\rho(h) = \rho^\chi_{\mu\nu}(h) 
:= \la \chi, \mu \vert \, h \, \vert \chi, \nu \ra\ .
\eqn\repmat
$$
The propagator (\pidef) is hence matrix-valued,
$K^{G/H}(q', q; T) = K^{G/H}_{\mu\nu}(q', q; T)$, with
indices given by the weight vectors belonging to the 
representation $\chi$.
The wave functions we consider thus take the values in the
linear space $V_\chi$ of the highest weight representation $\chi$.
The Peter-Weyl theorem then states that the set of the weight
factors (\repmat) provides a complete set of basis functions
for the wave functions on the group
manifold $H$, for which the orthonormality 
relations read
$$
\int_H dh \,  
\rho^\chi_{\mu\nu}(h)\, \rho^{\chi'*}_{\mu'\nu'}(h)
= \frac{1}{d_\chi}
\,\delta^{\chi\chi'}\,\delta_{\mu\mu'}\,\delta_{\nu\nu'}\ ,
\eqn\normbasis
$$
where $ d_{\chi} = \hbox{dim}\,V_\chi$ is 
the dimension of the representation $\chi$.
(Incidentally, we mention that the propagator
(\pidef) depends on the sections used.  For example,
if the initial  
point $q$ lies in the covering $D_\alpha$ whereas 
the final point $q'$ lies in the overlap $D_\beta \cap D_\gamma$ 
of the two
coverings, $D_\beta$ and $D_\gamma$, then 
the propagator
$K^{G/H}_{\mu\nu}(q', q; T)^{(\gamma\alpha)}$
defined using $\s_\gamma(q')$ and $\s_\alpha(q)$
is equal to 
$\rho_{\mu\lambda}^\chi(\s_\gamma^{-1}(q')\s_\beta(q'))
K^{G/H}_{\lambda\nu}(q', q; T)^{(\beta\alpha)}$ where the latter
propagator is defined
using $\s_\beta(q')$ and $\s_\alpha(q)$.)

We note in passing that, if the propagator on $G$ fulfills
the standard normalization condition, 
$\lim_{T \rightarrow 0} K^G(g', g; T) = \delta_G(g, g')$,
then  
the propagator on $G/H$ defined in (\pidef) satisfies
$$
\lim_{T \rightarrow 0} K^{G/H}_{\mu\nu}(q', q; T) =
\delta_{\mu\nu}\,\delta_Q(q, q')\ ,
\eqn\initial
$$
where
$\delta_G(g, g')$ and $\delta_Q(q, q')$ are delta-functions
on $G$ and $Q = G/H$, respectively.

In conclusion, we see 
that our guiding principle leads to the path-integral
formula (\pidef) for a homogeneous space 
possessing weight factors given by a unitary representation
of the isometry group $H$.  The significance of the obvious similarity
between the two path-integral formulae, (\pidef) and (\ldw),
will be discussed in Section 4.

\medskip
\noindent
{\bf 2.2. The holonomy factor}

Now we are in a position to derive the path-integral on $G/H$
possessing the holonomy factor from the definition
(2.5) which, from what we have found, becomes
$$
K_{\mu\nu}^{G/H}(q', q; T) 
  = \int_H dh \, \rho_{\mu\nu}^\chi(h) \,
    \int_{q}^{q'} \dq \, \int_e^h \dh \, e^{i S_G(\s(q)h)} \ .
\eqn\decpi
$$
Here we decomposed the path-integral measure $\dg = \dq\,\dh $ 
based on the decomposition (\candec)
intending to integrate out the
$H$ degrees of freedom to get an expression
purely in terms of the path-integration on $G/H$.
The decomposition (\candec) also induces the decomposition
of the action as
$$
S_G(\s(q)h) = S_{G/H}(q) + S'(q, h)\ .
\eqn\decaction
$$
The first term is the action on $G/H$,
$$
S_{G/H}(q) = \int_0^T dt
                \bigl(
                \frac{1}{2} || \dot q ||^2 - V ( \s(q) )
                \bigr) ,
\eqn\actiongoh
$$ 
which possesses the usual kinetic term of the particle moving under 
the induced metric of $G/H$ given by the norm $|| \cdot ||$ in $G/H$,  
{\it i.e.}, $|| \dot q ||^2 := \tr (\s^{-1}(q) \dot\s (q)\vert_\r )^2 
$ (we denote the decomposition of an element $X$ of the Lie algebra $\g$ 
of $G$ as $X = X\vert_\h + X\vert_\r$, where $\h$ is the Lie algebra
of $H$ while $\r$ is the orthogonal complement; see Appendix).
On the other hand, the second term in (\decaction),
$$
S'(q, h) = \int_0^T dt \, 
           \frac{1}{2}\tr \bigl(\dot h h^{-1} + A(q)\bigr)^2 \ ,
\eqn\saction
$$
contains the kinetic term for $h$ and the canonical connection
$A(q) := \s^{-1}(q) \dot\s (q) \vert_\h$ mentioned earlier.  

We here observe that, if we change the variables in the path-integral 
measure $\dh$ in (\decpi) as $h(t)_{\rm old} \rightarrow h(t)_{\rm 
new} := \tilde h^{-1}(t) h(t)_{\rm old}$ for $0 < t < T$ in such a way 
that the function $\tilde h(t)$ obeys the differential equation
$$
 \dot { \tilde  h}\,{\tilde h}^{-1} + A(q) = 0\ ,
\eqn\hcond
$$
then we can eliminate the canonical connection $A(q)$
from the action, thereby reducing 
$S'(q, h)$ to the action for the free particle
on $H$,
$$
S_H(h) = \int_0^T dt \, \frac{1}{2} || \dot h ||^2 \ .
\eqn\haction
$$
This, however, does not imply that the $q$-dependence
disappears entirely from the final result, because 
the change of variables causes a shift
in the boundary values of the path-integral in (\decpi) as
$\int_e^h \dh \rightarrow 
\int_{\tilde h^{-1}(0)}^{\tilde h^{-1}(T)\,h} \dh $.
But this $q$-dependence in the boundary values 
can also be eliminated by changing the variable in 
the integration $dh$ in (\decpi) analogously as
$h_{\rm old} \rightarrow h_{\rm new}
:= \tilde h^{-1}(T) h_{\rm old}$.
Choosing
the initial condition $\tilde h(0) = e$ for the solution 
of the equation (\hcond), we find 
after these successive changes of variables that the path-integral is 
decomposed into two parts; one over $G/H$ and the other over $H$,
$$
K_{\mu\nu}^{G/H}(q', q; T) = 
\int_{q}^{q'} \dq \, \rho_{\mu\eta}
      ^\chi(\tilde h(T))\, e^{i S_{G/H}(q)}\,\cdot\,
\int_H dh \, \rho_{\eta\nu}^\chi(h) \,\int_e^h \dh \, e^{i S_H(h)}\ .
\eqn\pitwo
$$

Let us first evaluate the second part in (\pitwo).  
Note that the Hamiltonian 
$\widehat H$ describing the free particle on the group manifold $H$
corresponding to (\haction) is just (half) 
the quadratic Casimir operator 
whose orthonormal eigenfunctions are given by the set $\{ 
\sqrt{d_\chi} \rho^{\chi *}_{\mu\nu}(h)\}$ in (\repmat).  Thus if we 
recall the standard formula that the path-integral $\int_x^{x'} {\cal 
D} x\, e^{iS(x)}$ can be expressed as 
$\langle x' \vert e^{-i\widehat H T} \vert 
x \rangle = \sum_n u_n(x') e^{-iE_nT} u^*_n(x)$ in terms
of the eigenfunctions $u_n(x) = \langle 
x \vert n \rangle$ of energy $E_n$, we find 
that the path-integral on $H$ in the second part reads
$$
\int_e^h \dh \, e^{i S_H(h)}
=  \sum_{\chi, \mu, \nu} \, d_{\chi}\, 
    {\rho^{\chi*}_{\mu \nu}}(h) \,
   e^{-{i\over 2}C_2^{\chi} T} \, \rho^{\chi}_{\mu \nu}(e)\ ,
\eqn\proph
$$
where $C_2^\chi$ is the value of the Casimir on the eigenfunction 
$\sqrt{d_\chi} \rho^{\chi *}_{\mu\nu}(h)$.  Then, with the orthonormality
relations (\normbasis), the second part is evaluated to be
$$
\int_H dh \, \rho^\chi_{\eta\nu}(h) \,  
  \sum_{\chi', \mu'} 
  \, d_{\chi'}\, \rho^{\chi'*}_{\mu' \mu'}(h) \, 
  e^{-{i\over 2}C_2^{\chi'} T}
= e^{-{i\over 2}C_2^{\chi} T}\, \delta_{\eta\nu}\ .
\eqn\normdet
$$

{}For the first part in (\pitwo) we only need to note 
that the solution of (\hcond) with the required initial
condition is given by the path-ordered product of 
the canonical connection, {\it i.e.}, the holonomy factor,
$$
\tilde h(T) = {\cal P}\, e^{- \int_0^T dt\, A}\ .
\eqn\holo
$$
Substituting (\normdet) and (\holo) 
into (\pitwo), we finally obtain
the desired expression,
$$
K_{\mu\nu}^{G/H}(q', q; T) =
e^{-{i\over 2}C_2^{\chi} T} \int_{q}^{q'} \dq \, 
\rho_{\mu\nu}^\chi\bigl({\cal P}\, e^{- \int_0^T dt\, A}\bigr) 
\, e^{i S_{G/H}(q)}\ ,
\eqn\pifinal
$$
which is the path-integral derived in [\LL] based
on the Trotter formula.  (The expression (\pifinal) has also been 
derived in our previous paper [\TT] using a stationary point
approximation; but here we have presented a better derivation
without any approximation.)

\medskip
\noindent
{\bf 2.3. Implications of the guiding principle}

Having derived the path-integral formula on $Q = G/H$
in the desired form (\pifinal), 
we now return to the path-integral (\pidef) which we started with 
and discuss what it actually implies.  For this purpose, it is 
instructive to consider the simple example $Q = S^1$.  According to 
the guiding principle, one has to lift it to a group manifold which is 
simply connected ($S^1$ is already a group manifold $S^1 = U(1)$ but 
it is not simply connected).  An obvious candidate is its universal 
covering space $\R$ with which $S^1 = \R / \Z$, where now $\R$ is 
regarded as the manifold of the group of translations on a line.  Then 
the above argument leading to the path-integral (\pifinal) applies 
formally to this case, even though the isometry group is discrete $H = 
\Z$ and hence one 
needs to replace, {\it e.g.}, the integration $\int 
dh$ by the summation $\sum_{n \in \Z}$.  But actually one does not 
need to go through the whole procedure to get the 
holonomy factor in (\pifinal), 
since the simplicity of the system allows for a direct derivation of 
the result almost from the outset once we follow the definition,
that is, the path-integral on $Q= S^1$ be given by
$$
K^{S^1} ( q' , q ; T ) =
  \sum_{ n = - \infty }^{ \infty } \, \rho(n) \,
        K^{\R} ( q' + 2 \pi n , q ; T ).
\eqn\pisone
$$
We have here used $q \in S^1 = [0, 2\pi)$ and assumed that the 
propagator on a line is invariant 
$ K^{\R} ( x'+ 2 \pi , x + 2 \pi ; T ) =
    K^{\R} ( x' , x ; T)$ for $x \in \R$
under the
translation of the group $H = \Z$,
which is ensured if the potential is periodic,
$V( x + 2 \pi) = V(x)$.

{}From the point of view
of $S^1$, it is clear that the summation over the final
points $q' + 2 \pi n$ 
in (\pisone) amounts to performing the summation
over all possible winding numbers
$n$ that the paths on $S^1$ can take during the transition
from $q$ to $q'$ on the circle.  The weight factor, which 
furnishes the representations of $\Z$, takes the form $\rho(n) = 
e^{in\theta}$ where $\theta \in [0, 2\pi)$ is the angle parameter 
specifying the representation of $\Z$.  Then, with $ A := \theta / ( 2 
\pi ) $, it is easy to rewrite the propagator (\pisone) into the form,
$$
        K^{S^1} ( q' , q ; T ) =
        e^{-i\frac{\theta}{2\pi}( q' - q )}\,
        \sum_{ n = - \infty }^{ \infty } \,
        \int_q^{ q' + 2 \pi n } \dq \,\,
        \hbox{exp} \biggl[
            i\int dt
             \bigl\{
                   \frac{1}{2} \dot q^2 - V (q)
                        + A \, \dot q \bigr\} \biggr] \ .
\eqn\pisofi
$$
We therefore see that the insertion
of the weight factor in this case $Q = S^1$
provides in effect
the minimal coupling with the vector potential $ A $.
Being constant, the vector potential has a vanishing curvature on
$S^1$ but the flux penetrating the circle is finite and proportional
to the parameter $\theta$.  Hence, its
physical consequence is analogous to that of 
the Aharonov-Bohm effect [\AB].

The appearance of the induced gauge potential $A$ is hence
well illustrated by the example $Q = S^1 = \R/\Z$, but unfortunately
another important element which in general appears in the 
path-integral on $Q = G/H$ is missing due to its simplicity.  The 
element missing is the matrix-valuedness of the propagator, which 
means that the Hilbert space consists of vector-valued wave 
functions taking values in $V_\chi$.  This implies that in addition to 
the degrees of freedom represented by the position of the particle on 
$G/H$, we have some finite degrees of freedom represented by the 
vectors, which we may regard as a generalized spin since they reduce to 
the conventional spin for $H = SU(2)$.  (The generalized spin 
does not arise for $S^1 = \R/\Z$ since the subgroup $\Z$ has only one 
dimensional irreducible representations.) These extra degrees of 
freedom are in fact necessary in summing up paths connecting $\s(q)$ 
and $\s(q')h$ in which the particle can move along the fibre $H$ as 
well in the base space $G/H$.  Since we have the kinetic energy 
term in $S_H$ in (\haction) (which is absent for $H = \Z$), in the 
generic case there exists a contribution from the integration along 
$H$.

The above feature seen in the formula (\pidef) may somewhat be 
counter-intuitive as a path-integral over the base space
$G/H$, but there is another, 
clearer characterization of the formula (\pidef) as follows.  
Suppose we quantize a system on $G$ which is invariant under
the action of $H$.  The symmetry then allows for considering
subspaces of the Hilbert space 
$L_2(G)$, which are classified according to the irreducible
representations $\chi$
of the symmetry group $H$.  The wave functions $\varphi(g)$
belonging to a subspace $\Gamma^\chi(G) \subset L_2(G)$
labeled by $\chi$ are hence $V_\chi$-valued
and characterized
by the {\it $\chi$-equivariance} under the $H$-action,
$\varphi(gh) = \rho^\chi(h^{-1})\varphi(g)$.  
Thanks to the symmetry, the propagation 
within the subspace $\Gamma^\chi(G)$, 
$$
(U^G(T)\varphi)_\mu(g')
:= \int dg \, K^{G}(g', g; T) \, \varphi_\mu(g)\ ,
\eqn\prpg
$$
is well-defined, {\it i.e.}, 
it furnishes a 
map $U^G(T): \, \Gamma^\chi(G) \rightarrow \Gamma^\chi(G) $.  Now, 
using (the pullback of) some section $\s(q)$, we may define an isomorphism 
$\Pi$ from $\Gamma^\chi(G)$ to the space of $V_\chi$-valued 
functions on $G/H$, denoted by $\Gamma^\chi(G/H)$, and its inverse 
$\Pi^{-1}$ by
$$
\Pi:\, \psi_\mu(q) := \varphi_\mu(\s(q))\ , \qquad
\Pi^{-1}: \, \varphi_\mu(g) 
    := \rho_{\mu\nu}^\chi\bigl(g^{-1}\s(\pi(g))\bigr) \, \psi_\nu(\pi(g))\ 
    , \eqn\no
$$
for $\varphi \in \Gamma^\chi(G)$ and $\psi \in \Gamma^\chi(G/H)$.  
Then, the propagation on $\Gamma^\chi(G/H)$,
$$
(U^{G/H}(T)\psi)_\mu(q')
:= \int dq \, K_{\mu\nu}^{G/H}(q', q; T) \, \psi_\nu(q)\ ,
\eqn\prpgoh
$$
may be defined 
with the help of the propagation on $\Gamma^\chi(G)$ as
$U^{G/H} := \Pi\circ U^G \circ\Pi^{-1}$.  Explicitly, it reads
$$
\eqalign{
(U^{G/H}(T)\psi)_\mu(q')
&= 
\int dg\, K^G(\s(q'), g) \, \rho_{\mu\nu}^\chi
\bigl(g^{-1}\s(\pi(g))\bigr) \, \psi_\nu(\pi(g)) \cr &= \int dq\,dh\, 
\rho_{\mu\nu}^\chi(h^{-1})\, K^G\bigl(\s(q')h^{-1}, \s(q)\bigr) 
\, \psi_\nu(q)\ , } \eqn\fwft
$$
where use has been made of the invariance (\hinv).
Comparing this to (\prpgoh) and noting 
the invariance $dh^{-1} = dh$ of the Haar measure, we find that
the propagator $ K^{G/H}(q', q; T)$ so defined reproduces
the formula (\pidef).  This shows that the three step
programme 
of quantization on $G/H$ stipulated by 
the guiding principle is actually equivalent to the above
procedure involving the projection from the space 
of the equivariant functions 
$\Gamma^\chi(G)$ to the space $\Gamma^\chi(G/H)$. 
The formula (\pidef) provides a concise expression of this
procedure in the path-integral framework.

\bigskip

\noindent \secno=3 \meqno=1

\centerline
{\bf 3. Description by the scalar-valued path-integral}
\medskip

The path-integral on a homogeneous space $Q$ derived
in Section 2 is generically matrix-valued.  However, in 
ref.[\MT] it was shown that one can also describe Mackey's quantization 
using a scalar-valued path-integral.  The key observation in [\MT] is that 
it is possible to realize the generalized spin taking its value in 
$V_\chi$ as the quantized version of the classical system of a 
coadjoint orbit of the group $H$.  In this section we provide a direct 
proof that the two path-integrals, matrix-valued and scalar-valued, 
are different expressions of the same quantization on $Q$.

\medskip
\noindent 
{\bf 3.1.  Scalar-valued vs vector-valued wave functions}

Before going into the proof, we recall first the description
of Mackey's inequivalent quantizations in terms of 
scalar-valued wave functions [\MT].  The wave functions that 
Mackey considered are vector-valued functions $\psi_\mu(q) \in 
\Gamma^\chi(G/H)$ characterized by the induced representation,
$$
(U_{\rm L}(g)\psi)_\mu(q)= \sum_\nu \rho^\chi_{\mu\nu}
\bigl(\s^{-1}(q) g \s(g^{-1}q)\bigr) \,\psi_\nu(g^{-1}q)\ ,
\eqn\induced
$$
under the left $G$-action $q \rightarrow g^{-1}q$
on $G/H$.
Our idea of constructing physically 
equivalent but scalar-valued
wave functions is that we start by scalar-valued
wave functions on $G$ and then find a
suitable reduction to wave functions on $G/H$ 
in such a way that the degrees of freedom
of $G$ be reduced to those of $G/H$ plus 
the generalized spin.  
The wave functions on $G$ are assumed to
transform under the $G$-action by the usual (left and right) regular 
representations,
$$
(U_{\rm L}(\tilde g)\psi)(g)=\psi(\tilde g^{-1}g)\ ,
\qquad
(U_{\rm R}(\tilde g)\psi)(g)=\psi(g\tilde g)\ ,
\eqn\reg
$$
for $\tilde g \in G$.
Thus the infinitesimal generators $\widehat L_m := \tr(\widehat L T_m)$, 
$\widehat R_m:= \tr (\widehat R T_m)$ associated with these left and 
right $G$-action corresponding to the basis $\{T_m\}$ of the Lie 
algebra $\g$ of $G$ fulfill the commutation relations,
$$
[\widehat R_m\,, \widehat R_n]= i f^l_{mn}\widehat R_l\,,
\qquad
[\widehat L_m\,, \widehat L_n] = i f^l_{mn} \widehat L_l\,,
\qquad
[\widehat R_m\,, \widehat L_n]= 0\,,
\eqn\qpbs
$$
where $f^l_{mn}$ are structure constants of the algebra $\g$.

To carry out the reduction we 
employ Dirac's approach [\Dirac] to quantizing a constrained
system, where a set of constraints are imposed as \lq physical state 
conditions' restricting the Hilbert space $L_2(G)$ to an appropriate 
subspace.  To choose the constraints in our case, 
we consider the set of 
operators,
$$
\widehat \phi_i  = \tr (\widehat R - K) T_i  \qquad
\hbox{for}\quad T_i \in \h \ , \eqn\newcons
$$
where $\{T_i\}$ is a basis in the Lie algebra $\h$ of the subgroup $H$,  
and $K \in \h$ is a constant element which will be specified below.  
For $K = 0$ the physical subspace, consisting of states satisfying 
$\hat \phi_i \psi_{\rm phy} (g) = 0$, reduces to the Hilbert space 
$L_2(G/H)$, because then the physical wave functions belonging to the 
subspace do not depend on the $H$ degrees of freedom $\psi_{\rm 
phy}(gh)= \psi_{\rm phy}(g)$ and can be regarded as wave functions on 
$Q = G/H$.  For $K \ne 0$, however, we find from the commutator
$$
[\widehat\phi_i\,, \widehat\phi_j] =
if^k_{ij}\,\widehat\phi_k+i\tr([T_i,T_j]K)\,,
\eqn\commut
$$
that $\{\widehat\phi_i\}$ in (\newcons) form a mixed (first and second class) 
set of operators and thus cannot be directly used to form a set of 
constraints to define the physical subspace.

To proceed, we have to first find a maximal set of first class components 
among the mixed set of operators, and to this purpose we consider the 
subalgebra ${\frak s}_K := {\rm Ker(ad}_K) \cap \h$ 
given by the kernel of the adjoint action of 
$K$ in $\h$.  For a generic 
$K$ ({\it i.e.}, if $K$ is a regular semisimple element in $\h$), the 
subalgebra ${\frak s}_K$ is the Cartan subalgebra of $\h$ 
containing $K$ [\Humphreys] (for $K$ non-generic, see [\MT]).  Choosing 
a basis $\{T_r\}$ in ${\frak s}_K$, we see that for any $T_j \in \h$ 
we have $\tr ([T_r, T_j] K)=0$ and hence the first class components in 
(\newcons) are
$$
\widehat\phi_r := \tr T_r (\widehat R - K) \qquad \hbox{for} \quad T_r \in 
{\frak s}_K \ .  \eqn\firstclass
$$
Conversely, from the semisimplicity of $\h$ it follows
that these $\widehat \phi_r$ form
the maximal set of the first class components in
(\newcons).
We now
need to find, from the operators in
(\newcons), the maximal subalgebra which can be used to define the physical
subspace.  For this, let us consider the complex extension $\h_{\rm c}$ of 
the algebra $\h$ and introduce a Chevalley basis $\{H_{\a_r}, 
E_{\pm\varphi}\}$ in $\h_{\rm c}$ (see Appendix) where we take $T_r := 
\frac{1}{i}H_{\a_r}$ ($r = 1, \ldots,{\rm rank}\,H$).  Then the 
maximal subalgebra of (\newcons) is given by the set $\{\widehat\phi_r 
,\widehat\phi_{\varphi}\}$, where
$$
\widehat\phi_{\varphi} = \tr E_{-\varphi}(\widehat R - K)\,
=\tr(E_{-\varphi}\widehat R),
\eqn\no
$$
for all positive roots $\varphi$. 
We then use these operators to define the physical states as
$$
\widehat\phi_r\, \psi_{\rm phy}(g) = 0 \qquad\hbox{and}\qquad
\widehat\phi_{\varphi}\, \psi_{\rm phy}(g) = 0\,.
\eqn\pscond
$$

If the constants $K_r = \tr(T_r K)$ take the integer values\note
{This integer-valuedness of the constant matrix 
$K$ can be derived from a consistency
condition in the path-integral [\letter, \MT]; see below.} 
corresponding to the highest weight $\chi$, {\it i.e.},
$$
K_r = 
\chi(H_{\alpha_r})\,,\qquad{\rm for}\quad 
r=1,\dots,\,{\rm rank\,}H\,,
\eqn\quantizek
$$
then, on account of the property
$$
\widehat{R}_r\rho^{\chi*}_{\mu\mu'}(h)=\mu'(H_{\alpha_r}) 
\rho^{\chi*}_{\mu\mu'}(h)\ , \eqn\property
$$ 
the physical states defined by (\pscond) have the solutions
$$
\psi_{\rm phy}(g)
= \psi_{\rm phy}(q,h)= \sum_\mu\psi_\mu(q) \rho^{\chi*}_{\mu\chi}(h)\ .
\eqn\physk
$$
The wave functions (\physk) provide
a description of Mackey's quantization in terms of scalar 
wave-functions.  Indeed, given a physical wave function (\physk), 
the coefficient functions in (\physk) are obtained as
$$
\psi_\mu(q)
= d_\chi\int_H dh\,
\psi_{\rm phy}(q,h)\rho^\chi_{\mu\chi}(h)\ , \eqn\coeff
$$
which reproduce exactly the induced representation (\induced) under the 
left-regular representation in (\reg) [\MT].

Once the constraint point of view for Mackey's quantization is 
established, one can write down the effective path-integral 
implementing the constraints in the standard manner.  Choosing 
gauge fixing conditions $\xi_s(g) = 0$ for the gauge 
symmetry generated by 
the maximal set of first class components (\firstclass), we find 
[\letter, \MT]:
$$
K_{\rm eff}(g', g; T) = {\cal N}\, \int_g^{g'} \dg \, \delta(\xi_r)\, 
\Delta_\xi \, e^{i S_{\rm eff}}\ , \eqn\spi
$$
where ${\cal N}$ is a 
formal normalization constant and 
$\Delta_\xi := {\rm det}\, \vert\{ \phi_r, \xi_{r'} \}\vert$ is the 
usual determinant 
factor necessary to render the path-integral gauge-independent.  
The 
effective action $S_{\rm eff}$ appearing in the path-integral (\spi) 
consists of three parts $S_{\rm eff} = S_{G/H} + S_{\rm coa} + S_{\rm 
int}$, where $S_{G/H}$ is the action on $G/H$ (\actiongoh) 
whereas the other two are given by
$$
S_{\rm coa} = - \int_0^T dt \, \tr (K h^{-1} \dot h)\ , \qquad 
S_{\rm int} = - \int_0^T dt \, \tr (h K h^{-1} A(q))\ . 
\eqn\coaint
$$
Here $S_{\rm coa}$ is
the action for the coadjoint orbit of $H$ passing through $K$,
which is the system of generalized spin in the sense 
that variables parametrizing 
the coadjoint orbit $S :=
h K h^{-1} $ form the Lie algebra $\h$ under Poisson
bracket (or commutator after quantization) [\AM, \Perelomov].
The action $S_{\rm int}$ then describes
the interation of the generalized spin 
with the canonical connection $A(q)$ in a way similar
to Wong's particle coupled to a Yang-Mills field [\Wong].
Finally we note that, for the path-integral (\spi) to be well-defined
under the presence of the gauge symmetry, 
the constant matrix $K$ must take the integer
values (\quantizek) [\letter, \MT], and that this is consistent with the
fact that quantization of coadjoint orbits 
leads to irreducible representations
of the group.

\medskip
\noindent 
{\bf 3.2. Scalar-valued path-integral}

We now show that the effective scalar-valued path-integral (\spi) found 
in the 
Dirac approach can directly be obtained from our matrix-valued 
path-integral with the holonomy factor (\pifinal).  
More explicitly, we shall evaluate the propagator $ K_{\rm phy}^{G/H}(g', 
g; T) $ for the scalar-valued physical wave-functions (\physk) defined 
in
$$
(U^{G/H}(T)\psi_{\rm phy})(g') := \int dg \, K_{\rm phy}^{G/H}(g', g; T) \, 
\psi_{\rm phy}(g)\ , 
\eqn\prongh
$$
and show that it coincides with the effective path-integral (\spi).

To this end, we first substitute the physical states in (\prongh) with the 
solutions (\physk) and use the propagator (\pifinal) for the 
coefficent vector-valued wave functions to get
$$
 K_{\rm phy}^{G/H}(g', g; T) = d_{\chi} e^{-{i\over 2}C_2^{\chi} T}
 \int_q^{q'} \dq\,
 \rho^\chi_{\chi\chi}
 \bigl({h'}^{-1}\,{\cal P}\, e^{- \int_0^T dt\, A}\, h \bigr) \, e^{i 
 S_{G/H}(q)}\ .  \eqn\prgha
$$
What remains to do is to rewrite the holonomy factor in (\prgha) 
in terms of a path-integral over $H$.  This is achieved if we recall the 
construction of the generalized coherent state 
path-integral of the group $H$ 
[\Klauder], where the key formula is the \lq resolution of unity' 
identity,
$$
d_{\chi}\int_H dh \, h\,\vert \chi,\nu \rangle \langle 
\chi,\nu \vert 
\, h^{-1} = 1\ .  \eqn\ru
$$
This identity (\ru) can be derived from the orthonormality
relations (\normbasis) by putting $\chi = \chi'$ and $\nu = \nu'$ 
using (\repmat).  Note that the integrand of (\ru) is constant 
along the Cartan subgroup $S_K := e^{{\frak s}_K}$ of $H$, and hence
the integration over $S_K$ 
yields merely the volume $V_{S_K} = \int_{S_K} dh$ of 
the group $S_K$. 
Hence, if we employ the same trick used
in the path-integral (\spi) to isolate the measure
for $S_K$ in the measure $dh$ for $H$, we may write (\ru) as
$$
\int \widetilde{dh} \, h\,\vert \chi,\nu \rangle \langle \chi,\nu 
\vert \, h^{-1} = 1\ , \eqn\res
$$
where $\widetilde{dh} := d_{\chi} V_{S_K}\,  dh \, 
\delta(\xi_r)\, \Delta_\xi$
with $\xi_r(h) = 0$ being conditions to fix the $S_K$ degrees of freedom,
now regarded as functions of $H$.

To build up a coherent state path-integration, we break, as usual, the time 
interval $[0, T]$ into $N+1$ short slices of duration $\Delta t = { T 
\over {N+1}}$ as $ 0 = t_0 < t_1 < \cdots < t_N < t_{N+1} = T $.  
Inserting the identity (\res) with $\nu = \chi$ at each slice of time 
in the holonomy factor in (\prgha) and setting $h(0) = h$ and $h(T) = h'$,
we find
$$
 \rho^\chi_{\chi\chi}
 \bigl({h'}^{-1}\,{\cal P}\, e^{- \int_0^T dt\, A}\, h \bigr) = 
 \int \prod_{n = 1}^N \widetilde{dh}(t_n) \, 
 \prod_{n = 0}^N \langle 
 \chi, \chi \vert \, h^{-1}(t_{n+1}) \, {\cal P}\, e^{- 
 \int_{t_n}^{t_{n+1}} dt\, A(t)}\, h (t_{n}) \, \vert \chi, \chi 
 \rangle\ .  \eqn\discret
$$
In the limit $N \rightarrow \infty$ we may take
$$
\eqalign{
\langle \chi, \chi \vert
\, h^{-1}(t_{n+1}) 
&
\, {\cal P}\,  
e^{- \int_{t_n}^{t_{n+1}} dt\, A(t)}\, \
h(t_{n}) \, \vert \chi, \chi \rangle \cr
& 
\simeq 
{\rm exp}\bigl(\, -\langle \chi, \chi \vert
h^{-1}(t_n) \dot h(t_n) + h^{-1}(t_n)A(q(t_n))h(t_n)
\vert \chi, \chi \rangle \Delta t \,  \bigr) \ .
}
\eqn\sim
$$
Using the identity
$\langle \chi, \chi \vert \, X \, \vert \chi, 
\chi \rangle = i \tr ( K X )$ for $X \in \h$  
given in the Appendix, we 
obtain the coherent state path-integral expession for the holonomy 
factor\note{%
The derivation of the path-integral formula for the 
holonomy factor (Wilson loop) is also discussed in [\DP] based on a similar 
but seemingly different method.}
$$
\eqalign{
 \rho^\chi_{\chi\chi} \bigl({h'}^{-1}\,
  & {\cal P}\, e^{- \int_0^T 
dt\, A}\, h \bigr) \cr 
&= \lim_{N\to\infty} \int \prod_{n = 1}^N 
\widetilde{dh}(t_n) \, e^{-i \sum_n {\rm tr}\, K (h^{-1}(t_n) \dot 
h(t_n) + h^{-1}(t_n)A(q(t_n))h(t_n))\, \Delta t} \cr 
&= \lim_{N\to\infty} d_\chi^{N} V_{S_K}^N
 \int_h^{h'} \dh \,\delta(\xi_r)\, \Delta_\xi \, 
e^{-i \int_0^T\, {\rm tr} (Kh^{-1}\dot h + hKh^{-1} A(q))}\ . 
} 
\eqn\holo
$$
Inserting (\holo) into (\prgha), we obtain
the scalar-valued 
path-integral, 
$$
K_{\rm phy}^{G/H}(g', g; T) =  
{\cal N}\, \int_q^{q'} \dq   \int_h^{h'} \dh 
\, \delta(\xi_r)\, \Delta_\xi \, e^{i 
S_{G/H}(q) - i\int_0^T\, {\rm tr} (Kh^{-1}\dot h + hKh^{-1} A(q))}\ ,  
\eqn\finalsvpi
$$
with ${\cal N} = e^{-{i\over 2}C_2^{\chi} T}
\lim_{N\to\infty} d_\chi^{N+1} V_{S_K}^N$,   
which agrees precisely
with the effective path-integral\note{%
For completeness we mention that in [\letter, \MT] the factor
$e^{-{i\over 2}C_2^{\chi} T}$ is dropped during the path-integral
reduction on the ground that 
it merely gives a constant shift in the effective
Lagrangian.}
$K_{\rm eff}^{G/H}(g', g; T)$ in (\spi).
Although, as is well-known [\Klauder], derivation of path-integrals 
using the (generalized) coherent states remains purely formal,
the above argument provides another confirmation (along with the 
argument by the induced representations 
given in Section 3.1) in the path-integral framework
that the constraint
point of view on Mackey's quantization using Dirac's approach
is in fact valid.

\bigskip

\noindent \secno=4 \meqno=1

\centerline
{\bf 4.  Conclusion and discussion}
\medskip

We have seen in this paper that the \lq lift and then project' principle
leads to a path-integral which reproduces correctly the two important 
aspects of the quantum theory on a homogeneous space $G/H$, namely,
the inequivalent quantizations and the induced gauge field.
The path-integral formula on $G/H$ derived from the principle
is given by the path-integral on $G$ with the $H$ degrees of freedom
smeared out with weight factors.  Consistency (composition law) 
requires the weight factors
to form a representation of $H$, rendering the entire formula
analogous to the familiar one for a multiply connected space
$\bar Q/\Gamma$ in [\LDW, \Dowker].         
We also derived the holonomy factor in the path-integral after 
isolating the integration on $G/H$ from that
on $H$, and this derivation indicates that 
the induced gauge field that appears in the
holonomy factor is a direct consequence of the weight factors, that is,
the essential ingredient of the inequivalent quantizations.  The scalar-valued
path-integral obtained in the effective Dirac approach [\letter, \MT] 
is then derived from our matrix-valued form using the coherent 
state path-integral
technique.

As mentioned in the Introduction, the fact that at least for 
the two types of spaces, $Q = G/H$ and $Q = \bar Q/\Gamma$,  quantum theory is
characterized by the irreducible unitary representations of the isometry
group, $H$ or $\Gamma$, and not just those of the fundamental group of the space
$\pi_1(Q)$, suggests that the path-integral formula 
(\pidef) presented here is a
generalization of the formula (\ldw) for a multiply connected space.
Once this interpretation is accepted, it is perhaps natural
to extend the path-integral approach further to inhomogeneous spaces.
In fact, inhomogeneous spaces often arise in physics,
with the one most frequently discussed being
a Riemann surface with higher genus.
A possible extension to inhomogeneous spaces 
is the following\note{%
This type of construction has already been
considered in [\Montgomery]
in investigating geometric properties of induced gauge fields
of deformable bodies, and has also been taken up recently in [\Robson] 
in an attempt to the extension based on the method of 
geometric quantization.} .
Let $ P $ be a Riemannian manifold with a metric $ g_P $ and
let a Lie group $ H $ act on $ P $ freely and isometrically.
The manifold $ Q = P/H $ then admits an induced metric $ g_Q $,
with which the projection $ \pi : P \to Q $
defines a principal bundle and a Riemannian submersion.
Now, given a propagator on $ P $ which is $ H $-invariant
$ K_P ( p'h, ph; t ) = K_P ( p', p; t ) $, 
a straightforward extension of our formulation of the path-integral
to such inhomogeneous spaces
is possible by defining 
the propagator on $Q$ as
$$
        K^Q( p', p; T )
        =
        \int_H dh \, \rho^\chi(h)\, K^P ( p'h, p; T ),
\eqn\extension
$$
which acts on $\chi $-equivariant functions $ f : P \to V_\chi $;
$ f ( p h ) = \rho^\chi ( h )^{-1} f ( p ) $.

However, the biggest problem with this extension is that 
there is no guarantee of having
a uniquely defined propagator on $P$ for describing the  
transition amplitude, because we do not know how to
quantize on $P$ in the first place.  In fact, the basic reason why 
we lift our system to the universal covering space $\bar Q$
in the case of a multiply connected space, or to 
a group manifold $G$ in the case of a homogeneous
space, is that we know in principle how to quantize on these spaces, and
this actually limits the scope of extension of quantization irrespective
of the approach employed.   Another problem of the extension
to inhomogeneous spaces is that we seem to lose any control
on the form of the induced connection.  This is so because,
when the base space $ ( M, g_M ) $ is fixed,
inequivalent quantizations depend on the choice of 
the principal bundle $P$, 
the lifted metric $g_P$ and the representation $\rho^\chi$,
which amounts to an introduction of an arbitrary connection.
This problem does not arise for a homogeneous space, where
there exists a criterion to choose a specific connection, since 
the invariance under the $ G $-action determines
both $ g_{G/H} $ and $g_G$ uniquely, leading to the $G$-invariant
canonical connection [\LMT].  The existence of such a transitive 
$G$-action is not only crucial for specifying the connection but
also important for the existence of momentum, because 
a self-adjoint momentum operator can then be defined globally as
a generator of the transitive action, which in turn is important
in defining a self-adjoint Hamiltonian to 
provide a unitary time evolution of the system.

If, on the other hand, we have some means to quantize on $P$,
then the formula (\extension) may provide a basis not just for quantizing
on $Q = P/H$ but for extending even further to field theory, such as
the Yang-Mills theory whose configuration space is  
${\cal A}/{\cal G}$, where ${\cal A}$ is the space of all gauge potentials
and ${\cal G}$ the group of gauge transformations.
Being an affine space, the space ${\cal A}$ seems to admit
a conventional quantization and, accordingly, may ensure
the extension of our path-integral formula, too.
We will not discuss
here the outcome of the application of the formula, but a preliminary 
investigation of the Yang-Mills
theory shows that there exist various quantum effects, in addition to
the celebrated 
$\theta$-vacuum structure which is the effect of $\pi_1({\cal A}/{\cal G}) = \Z$,
which we have not uncovered yet.  We hope to report the full detail
of the investigation elsewhere in the near future.

\bigskip
\noindent
{\bf Acknowledgements:}  
S.T. wishes to thank T.~Iwai and Y.~Uwano for their support in this
work, and I.T. is grateful to L.~Feh\'er, 
D.~McMullan and A.~Wipf for helpful discussions.
This work is supported in part by the Grant-in-Aid for Scientific
Research from the Ministry of Education, Science and Culture 
(No.07804015 and No.08740200).

\bigskip

\secno=0\appno=1\meqno=1

\noindent
{\bf Appendix. Conventions}
\medskip

In this appendix we shall provide our conventions 
on Lie groups/algebras together with some basic
facts on coset spaces used in the text (see, for example 
[\Humphreys, \Barut]).

Let G be a semisimple group and $\g$ its Lie algebra.
In the Lie algebra 
$\g$ (or in the complex extension $\g_{\rm c}$ of $\g$)
one can choose the
Chevalley basis 
$\{H_\alpha, E_{\pm\varphi} \}$ where $\alpha$
are simple roots and $\varphi$ are positive roots.
The basis satisfies the relations
$$
[E_\alpha, E_{-\alpha}] = H_\alpha\,, \qquad 
[H_\alpha, E_\beta] =
K_{\beta\alpha} E_\beta\,,
\eqn\kw
$$
for simple roots $\alpha$, $\beta$, and $K_{\beta\alpha} =
\beta(H_\alpha) =
{{2\beta\cdot\alpha}\over{\vert\alpha\vert^2}}$ is the Cartan \
matrix.
To every dominant weight $\chi$
there exists an irreducible representation
--- highest weight
representation --- of $\g$ in which the Cartan elements 
$H_\alpha$
are diagonal; in particular,
on the states $\ket {\chi, \mu}$
specified by the weights $\mu$ connected to
the dominant weight $\chi$ (identified as the highest weight
in the representation) the eigenvalues are all integer:
$$
H_\alpha \ket {\chi, \mu} = \mu(H_\alpha) \, \ket{\chi,  \mu} \
\,, \qquad
{\rm with}\qquad
\mu(H_\alpha) = {{2\mu\cdot\alpha}\over{\vert\alpha\vert^2}}
              \in \Z\,.
\eqn\hwr
$$
On account of this, we can use the dominant weight $\chi$ (or \
the set of
integers $\chi(H_{\alpha_r})$ for $r = 1, \ldots, {\rm rank}\,G$)
to label the irreducible
representation.  
(To avoid any confusion, we note that in the text 
the Chevalley basis and the highest weight representations are
considered for the subgroup $H$ of $G$, not for $G$.)

Given a closed subgroup $H$ of $G$
with its Lie algebra $\h$,
we take an orthogonal decomposition of $\g$,
$$
\g=\h\oplus\r\,,
\eqn\od
$$
where $\r=\h^\perp$ is the orthogonal complement of $\h$ to 
$g$
with respect to the innerproduct $(\h, \r) = 0$ which is defined
by the trace in an irreducible 
representation $\rho$;  
$(X, Y) = \tr (\rho(X)\, \rho(Y))$ for $X$, $Y \in \g$.
(We often omit the $\rho$ in the trace and write it simply as
$\tr(X\,Y)$.)
The decomposition (\od) is, in fact, reductive, {\it i.e.,}
$[\h,\r]\subset\r$,
because $[\h,\h]\subset\h$ and the
orthogonality imply 
$0 = ([\h,\h], \r) = (\h, [\h,\r])$.
The decomposition of an element $X \in \g$ according to (\od)
is written as $X = X\vert_{\h} + X\vert_{\r}$.
We shall denote bases of the spaces by
$$
\eqalign{
\g &= \hbox{span} \{T_m\}\,, \cr
\h &= \hbox{span} \{T_i\}\,, \cr
\r &= \hbox{span} \{T_a\}\,, \cr
}
\qquad
\eqalign{
m &= 1, \ldots, \dim{ G}\,, \cr
i &= 1, \ldots, \dim{ H}\,, \cr
a &= 1, \ldots, \dim{ (G/H)}\,. \cr
}
\eqn\conv
$$
Using a properly normalized trace one has
$(T_m, T_n) = \delta_{mn}$.

Choosing in particular the Chevalley basis 
$\{ H_\alpha, E_{\pm\varphi} \}$ 
for our basis in $\g$, we can expand any anti-Hermitian element,
$X \in \g$ with $X^\dagger = - X$, as
$$
X = \sum_\alpha X_\alpha \, {1\over i} H_\alpha + 
\sum_{\pm \varphi} X_{\pm \varphi}\, E_{\pm \varphi} \ ,
\eqn\expansion
$$
where $X_\alpha$, $X_{\pm \varphi}$ are the coefficients
of the expansion.  
Let $K$
be the element belonging to the Cartan subalgebra of $\g$
carrying the label of the representation $\chi$ in the 
coefficients such that $({1\over i} H_\alpha, K) = \chi(H_\alpha)$.  
Then we have the identity
$$
\bra{\chi,\chi}\, X \, \ket{\chi,\chi} = i \, (K, X)\ .
\eqn\identrace
$$

We can furnish
a Riemannian metric on the coset space $G/H$
as follows.  
Decompose first $g \in$ G as $g = \s \, h$
where $\s \in {G}$ is a (local) section and $h \in {H}$.
Then form the Maurer-Cartan 
1-form $\s^{-1}d\s$ and split it
as
$\s^{-1} d\s = \s^{-1}d\s \vert_\h + \s^{-1}d\s \vert_\r $, where
the first part $\s^{-1}d\s \vert_\h $ leads to the canonical connection
$A$ mentioned in the text.  
The second part $e := \s^{-1}d\s \vert_\r$, on the other hand, 
is regarded as a vielbein, from which 
the metric on $G/H$ is defined by
$$
ds^2 = g_{\alpha\beta} \, dq^\alpha \otimes dq^\beta
:= \tr (e \otimes e)\,,
\eqn\metgh
$$
where $\{q^\alpha \}$, $\alpha = 1, \ldots, n$
($n = \dim(G/H)$), is a set of local coordinates on
$G/H$.

\vfill\eject

  \vfill\eject\immediate\closeout\reffile
  \centerline{{\bf References}}\bigskip\frenchspacing%
  \input refs.tmp\vfill\eject\nonfrenchspacing

\bye